\documentclass[proof]{WileyASNA-v1}

\articletype{Article Type}%

\received{}
\revised{}
\accepted{}

\begin{document}

\title{Non-symmetrical sparking may hint ``zits'' on a pulsar surface}

\author[aff1]{Zhengli Wang}

\address[aff1]{Guangxi Key Laboratory for Relativistic Astrophysics, School of Physical Science and Technology,
Guangxi University, Nanning 530004, China}

\author[aff2,aff3]{Jiguang Lu}

\address[aff2]{National Astronomical Observatories, Chinese Academy of Sciences, Beijing 100012, China}
\address[aff3]{Guizhou Radio Astronomical Observatory, Guiyang 550025, China}

\author[aff2]{Jingchen Jiang}

\author[aff4,aff5]{Shunshun Cao}

\address[aff4]{School of Physics and State Key Laboratory of Nuclear Physics and Technology, Peking University, Beijing 100871, China}

\address[aff5]{Kavli Institute for Astronomy and Astrophysics, Peking University, Beijing 100871, China}

\author[aff6]{Weiyang Wang}

\address[aff6]{School of Astronomy and Space Science, University of Chinese Academy of Sciences, Beijing 100049, China}

\author[aff1]{Enwei Liang}

\author[aff4,aff5]{Renxin Xu*}

\corres{*Renxin Xu, School of Physics, Peking University, Beijing 100871, China. \\
\email{r.x.xu@pku.edu.cn}}


%
%
%
%
%
%
\abstract{
Pulsar electrodynamics could be relevant to the physics of stellar surface, which remains poorly understood for more than half a centenary and is difficult to probe due to the absence of direct and clear observational evidence.
Nevertheless, highly-sensitive telescopes (e.g., China's Five-hundred-meter Aperture Spherical radio Telescope, FAST) may play an essential role to solve the problem since the predicted surface condition would have quite different characteristics in some models of pulsar structure, especially after the establishment of the standard model of particle physics.
For instance, small hills (or ``zits'') may exist on solid strangeon star surface with rigidity, preferential discharge, i.e., gap sparking, may occur around the hills in the polar cap region.
In this work, with the 110-min polarization observation of PSR B0950$+$08 targeted by FAST, we report that the gap sparking is significantly non-symmetrical to the meridian plane on which the rotational and magnetic axes lie.
It is then speculated that this asymmetry could be the result of preferential sparking around zits which might rise randomly on pulsar surface.
%
%
%
%
%
%
%
Some polarization features of both single pulses and the mean pulse, as well as the cross-correlation function of different emission regions, have also been presented.
}


\maketitle


\section{Introduction}

The underlying radiation mechanism of radio pulsars is yet not understood though the discovery of the pulsar has been more than a half century~\citep[e.g.,][]{2017JPhCS.932a2001M, 2018PhyU...61..353B}. 
With highly-sensitive telescopes, detailed observations of radio pulsars may provide an opportunity to understand these issues.
%
PSR B0950$+$08 is located nearby, with a distance of 0.26\,kpc only~\citep{2005MNRAS.364.1397J}. Several research groups have been focused on the radiation characteristics of this pulsar~\citep[e.g.,][]{1981ApJ...249..241H,1984ApJS...55..247S,2001ApJ...553..341E,2023arXiv230807691W}.
~\cite{1981ApJ...249..241H} presented the interpulse emission features from PSR B0950$+$08 and pointed out firstly that the radiation of this pulsar occupied extremely wide pulse phase. Its radio emission duty cycle is about 0.83.
~\cite{2022A&A...658A.143B} reported the emission properties of individual pulses of this pulsar based on dual-frequency (55\,MHz and 1.4\,GHz) observations. They found that the fluctuation emission in a single pulse over a short time scale (e.g. $\mu$s) is due to the effect of the diffractive scintillation.
~\cite{2022MNRAS.517.5560W} published the radiation characteristics of PSR B0950$+$08, using the Five-hundred-meter Aperture Spherical radio Telescope (FAST). They found that the radio signal of this pulsar occupies the whole pulse phase based on the long integration of 160-min observation.

\par It is worth noting that pulsars' polarization properties are closely related to the radiation geometry and even to the pulsar magnetosphere~\citep[e.g.,][]{1984ApJS...55..247S,2001ApJ...553..341E,2020ApJ...890..151R,2023arXiv230807691W}.
~\cite{1984ApJS...55..247S} published single pulse polarization observation of PSR B0950+08 at 1404\,MHz with the Arecibo radio telescope. They concluded that the linear polarization emission feature in the main pulse displays the depolarization, and explained this phenomenon as the underlying of the emission consisting of roughly equal contributions from the two modes simultaneously.
~\cite{2001ApJ...553..341E}studied the emission beam geometry of PSR B0950$+$08 by using the classical rotating vector model (RVM)~\citep{1969ApL.....3..225R} to fit the average polarization position angle (PPA) of this pulsar.
After taking the PPA of the ``bridge'' component emission across a wide longitude range into account, they concluded that this pulsar is an orthogonal rotator with the main and interpulse radiation emitted from opposite magnetic poles.
Recently,~\cite{2023arXiv230807691W} reported the result of the accurate polarization observation of PSR B0950$+$08 based on FAST radio telescope. They found that the distribution of the sparks of this pulsar is non-symmetrical around the magnetic pole by mapping the sparks onto the polar cap on the pulsar surface.
Besides the puzzling magnetospheric activity, to be related closely to the nature of pulsar surface, understanding pulsar's inner structure (i.e., the equation of state
of supra-nuclear matter at low temperature) should be more
challenging in today’s physics and astrophysics.
As for the state of dense supra-nuclear matter, {\it neutronization} was proposed for ``gigantic nudelus'' more then 90 years ago~\citep{Landau1932}, but {\it strangonization} has also been conjectured~\citep{Xu2003,2017JPhCS.861a2027X} after the establishment of the standard model of particle physics.
The latter is the focus of this paper, and we are investigating the observational consequence of preferential discharges for PSR B0950+08 as an example, since small hills would grow naturally on the surface of strangeon matter with rigidity, and sparking may occur preferentially around the hills.

\par In this work, we report the studies on emission feature of PSR B0950$+$08 based on the FAST telescope, showing more observational features, in order to understand the sparking behaviors and their relevance to pulsar's surface condition.
We summarize this observation and data processing in Section 2.
The results are given in Section 3.
%
%
Conclusions and discussions are presented in Section 4.

\begin{figure}[h!]
\centering
\includegraphics[scale=0.55]{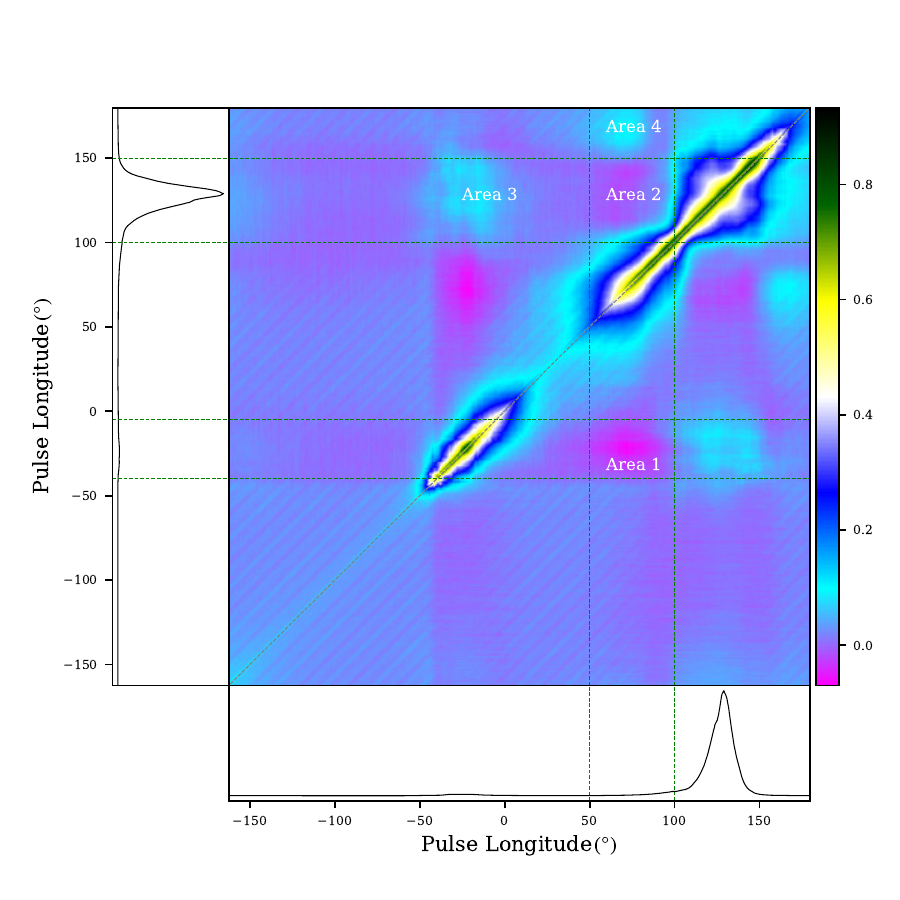}
\caption{The cross-correlation function map of the whole pulse phase. The region between two horizontal dashed lines (from $-45^{\circ}$ to $-10^{\circ}$) corresponds to the interpulse, and another region between two horizontal dashed lines (from $100^{\circ}$ to $150^{\circ}$) represents the main pulse. The correlation coefficient is indicated by the color. The two vertical dashed lines (from $50^{\circ}$ to $100^{\circ}$) are targeted to describe the precursor component of the main pulse. Magnetic poles are located at $\phi = 0^{\circ}$ and $\phi = \pm 180^{\circ}$.
} %
\label{ccf}
\end{figure}

\section{Observation and data processing}

PSR B0950$+$08 was observed with the FAST radio telescope on MJD 59820 (2022 August 29). The observation of full Stokes parameters was carried out. This polarization observation was arranged with the 19-beam receiver system of FAST over 110-min integrations~\citep{2020RAA....20...64J,2019SCPMA..6259502J}. More than 26000 individual pulses were obtained from this pulsars.

\par The \textsc{dspsr} software package~\citep{2011PASA...28....1V}was adopted in the data processing. The raw data was resampled by the \textsc{dspsr} software package, and the sampling time was around 0.25\,ms. In addition, radio frequency interference (RFI) is eliminated using the time-frequency dynamic spectrum.

\begin{figure}[h!]
\centering
\includegraphics[scale=0.415]{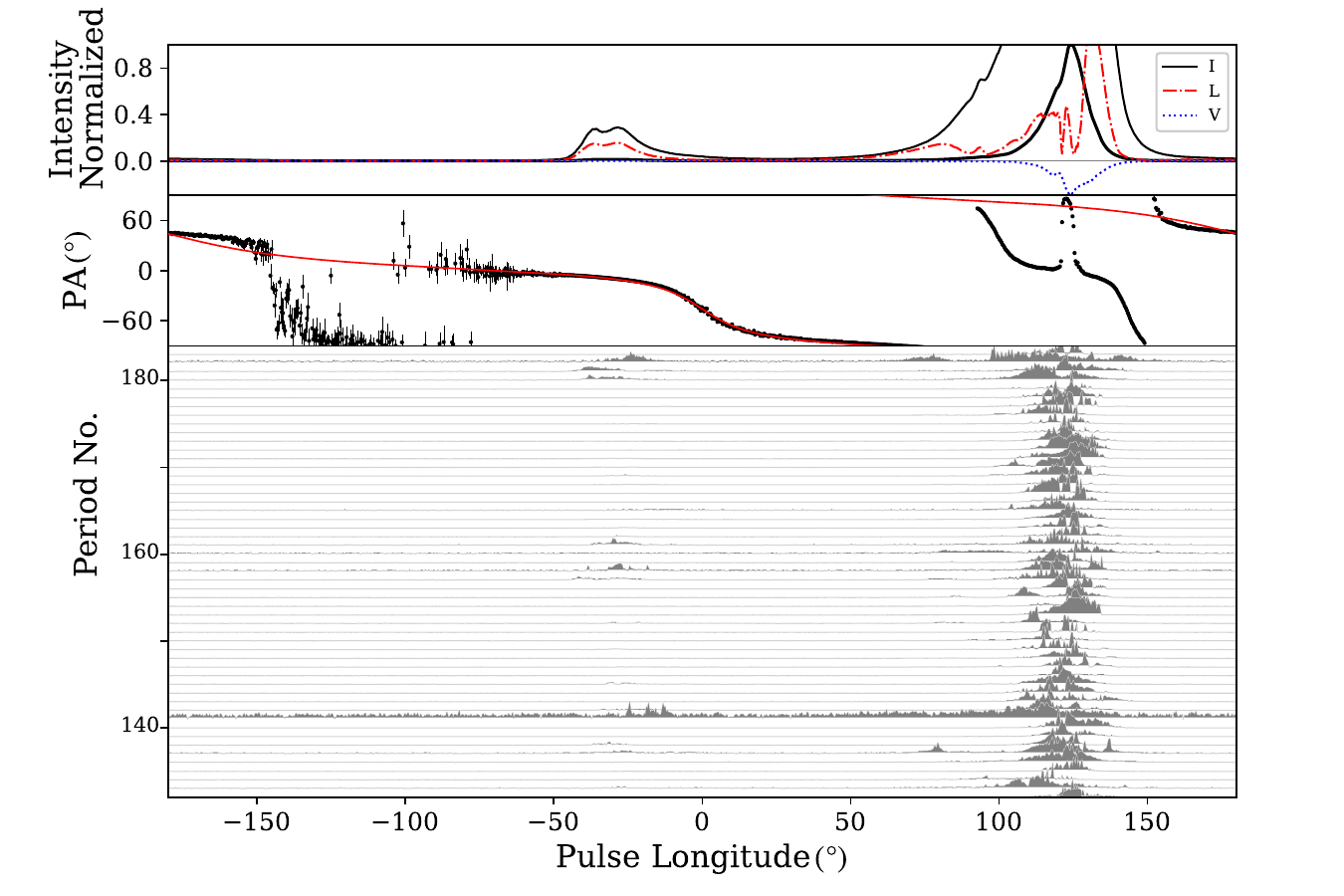}
\caption{The emission state of PSR B0950$+$08. The top panel depicts total intensity in thick black; then the total intensity (thin black), the total linear polarization (dashed red) and circular polarization (dotted blue) are plotted at $20 \times $ to show the interpulse and the polarization intensities in the main pulse clearly. The intensity is scaled with the radio emission peak. The middle panel describes the measured polarization position angle (PPA) and its best RVM fit (red). It gives the inclined angle $\alpha$ and the impact angle $\beta$ are $100.5^{\circ}$ and $-33.2^{\circ}$, respectively. The steepest gradient of the RVM curve has been centered. The radio emission behavior of the single pulse is shown in the bottom panel. Note that more peak structures are visible in some periods (i.e., Period Nos. 137 and 182).
} %
\label{Fp}
\end{figure}

\section{Results}

\subsection{The underlying physical mechanism over different pulse longitudes}

The physical process in the magnetosphere strongly affect the emission features of radio pulsars. To unravel the underlying physical processes of this pulsar in different emission regions, the cross-correlation function (CCF) as a function of the pulse longitude is analyzed. As Figure~\ref{ccf} shows, the map is divided into different areas according to the correlation between different emission regions in order to reveal the underlying physics. 
\par The ``Area 3'' corresponds to the correlation between the interpulse and main pulse, showing that the emission features of the interpulse and main pulse have a weak positive correlation. This feature means that the physical processes that respectively dominates the two regions are similar. 
The ``Area 1'' represents the correlation between the precursor component emission of the main pulse and the interpulse, and the ``Area 2'' reveals the correlation between the precursor component emission of the main pulse and the main pulse. Both of them show a weak negative correlation, implying that the physical mechanism is opposite to that of the interpulse and main pulse.
The ``Area 4'' depicts the correlation between the precursor and the postcursor component emission regions of the main pulse. It is worth noticing that the emission correlation of two weak emission regions (i.e., the precursor and the postcursor components of the main pulse) has similar properties. Furthermore, the emission correlation feature implies that a similar underlying physical mechanism dominates the two weak and strong emission regions (i.e., the interpulse and main pulse). 

\subsection{The emission properties of individual pulses}

\par The emission state over a short time scale may reveal the relevant radio emission phenomena (e.g., interstellar scintillation) along the path of the light of sight and even unravel the distribution of the sparks on the polar cap surface. 

\par To further clarify the sparks on the polar cap surface, the emission state of the individual pulses is analyzed. As Figure~\ref{Fp}shows, the emission state of the individual pulse exhibits random properties, particularly in the main pulse. Note that the radio signal of the individual pulse (period No. 141) occupies an extremely wide pulse longitude. Moreover, the pulse profiles of the two individual pulses (period Nos. 137 and 182) display more peak structures.
Certainly, these single pulses reflect a complex dynamics of gap sparking, and this emission feature may imply the different locations of small mountains (or ``zits'') on the stellar surface.
The different emission intensities would be interpreted as the different preferential discharges coming down to the heights of the mountains. This phenomenon is similar for many other pulses: more peak structures exhibit over a wide pulse longitude for a rotation period. Furthermore, these different peaks of small mountains may be located non-symmetrically on the stellar surface.

Other observational consequences of small hills inside polar cap may includes the strong and weak individual pulses of  pulsar B2111+46~\citep{2023NatAs...7.1235C} and the unusual arc-like structure of the bright pulsar PSR B0329+54~\citep{2007MNRAS.379..932M},  or the distinct core-weak patterns~\citep{2023MNRAS.520.4173W}.
Certainly, further studies of radio singe pulses with FAST are really encouraged to find solid evidence for zits on face.

\subsection{The pattern of the sparks on the polar cap surface}

\par
\begin{figure*}
\centering
\includegraphics[scale=0.50]{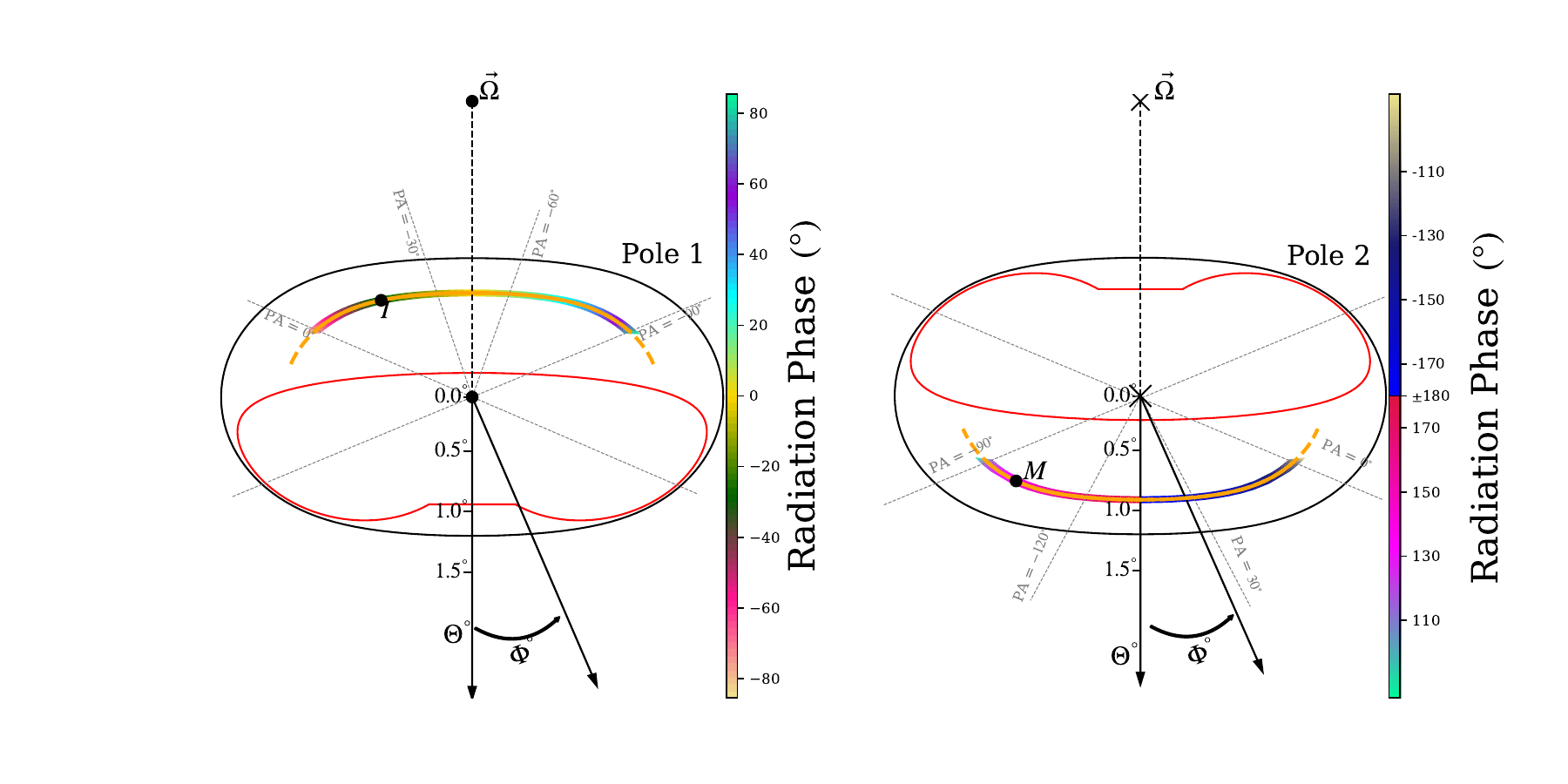}
\caption{The polar cap of PSR B0950$+$08 is plotted assuming a geometry of $\alpha = 100.5^{\circ}$ and $\beta = -33.2^{\circ}$ under the dipole field configuration.
The footprints of the last closed field lines are in solid black lines.
The footprints of the critical field lines (the field lines are perpendicular to the rotation axis at the light cylinder, i.e., $\mathbf{\Omega} \cdot \mathbf{B}$ = 0) correspond to the red line.
Without loss of the generality, we assume that the emission of this pulsar comes from the magnetic field lines whose footprints are the dashed orange curves.
The sparking trajectories are plotted by colors under the framework of the dipole field.
To further unravel the emission geometry of this pulsar, the PA is also included in the plot (the dashed grey lines).
The angles, $\Theta$ and $\Phi$, are for the polar and the azimuthal angles around the magnetic poles, respectively.
} %
\label{mapping}
\end{figure*}

 To further understand the emission property of this pulsar, the pattern of the sparks on the polar cap surface is analyzed by assuming the dipole field based on the 110-min polarization observation. The sparking pattern is shown in Figure~\ref{mapping}. Without loss of generality, we assume the emission from the magnetic field lines whose footprints concentrate on the three-quarter distance between the magnetic pole and the last closed field lines (i.e., the orange dashed lines). 

According to the relation between the viewing angle ($\zeta = \alpha + \beta$) and the emission point, one has
\begin{equation}
    \zeta = \cos^{-1} (\mathbf{\hat{B}_z}/ \sqrt{\mathbf{\hat{B}_x}^2 + \mathbf{\hat{B}_y}^2 + \mathbf{\hat{B}_z}^2}),
    \label{eq1}
\end{equation}
where $\mathbf{\hat{B}} = \mathbf{\hat{B}_x} + \mathbf{\hat{B}_y} + \mathbf{\hat{B}_z}$ is the unit vector of the magnetic field at the emission point. 
The solid curve refers to the footprints of the magnetic field lines that can be tangent to the light of sight within the light cylinder.
It is evident that the emission of the range $-85^{\circ}$ to $85^{\circ}$ is from the magnetic Pole 1, while that of the remaining $52 \%$ from its opposite magnetic pole.
The labels ``$I$'' and ``$M$'' correspond to the sparking locations for the interpulse and main pulse, respectively,  implying that those two regions of the preferential discharge on this pulsar's surface (i.e., for the interpulse and the main pulse) are far away from its magnetic pole.


\subsection{The distribution of the polarization emission orientation on a Poincaré sphere}
\par
\begin{figure*}
    \centering
    \begin{minipage}[t]{1\linewidth}
        \centering
        \includegraphics[width = 4.5in]{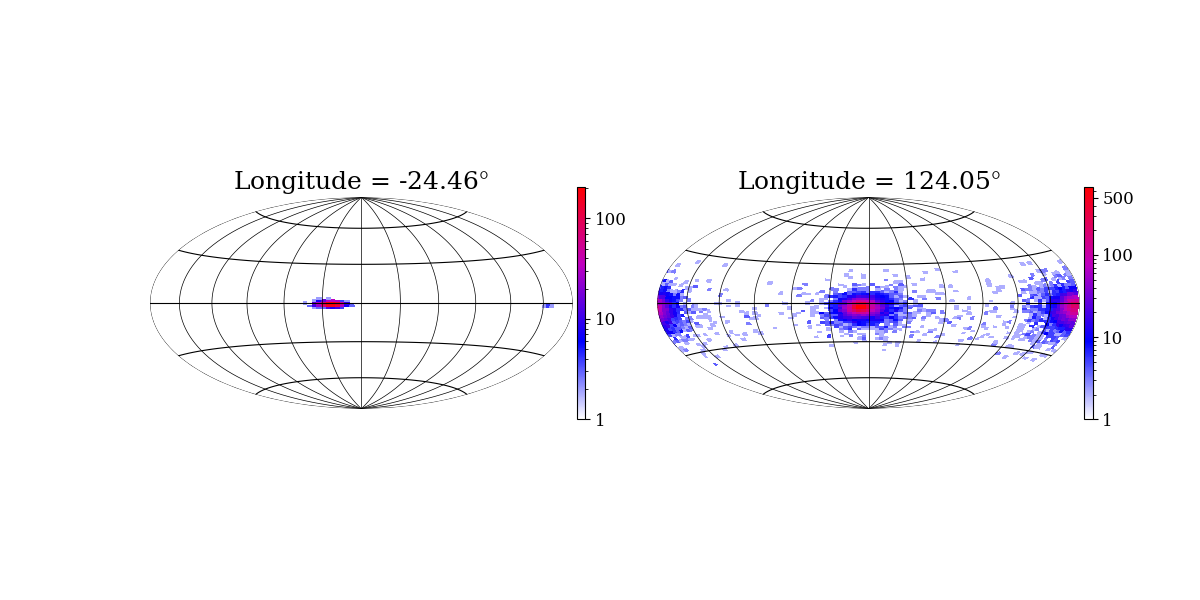}
        \centerline{(a)}
    \end{minipage}
    \begin{minipage}[t]{1\linewidth}
        \centering
        \includegraphics[width = 4.5in]{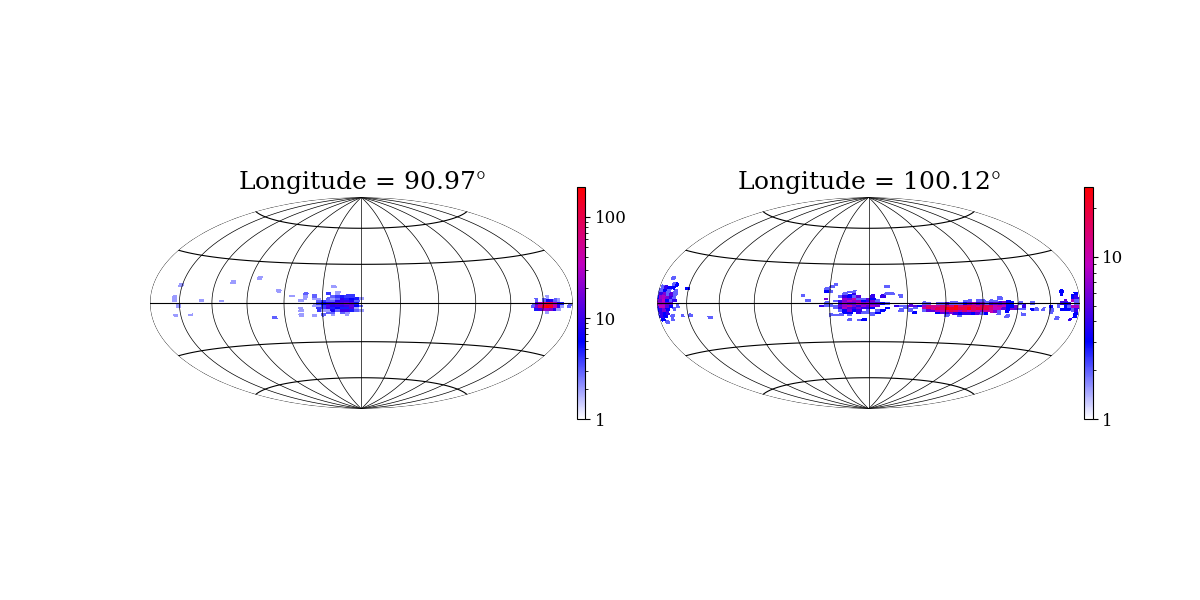}
        \centerline{(b)}
    \end{minipage}
    \begin{minipage}[t]{1\linewidth}
        \centering
        \includegraphics[width = 4.5in]{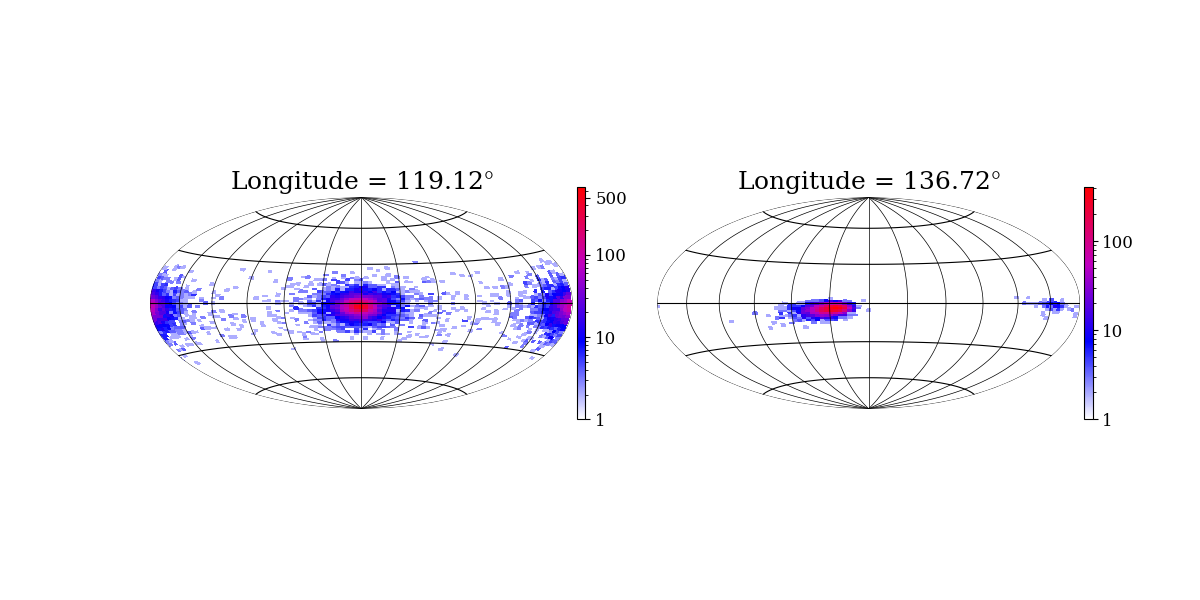}
        \centerline{(c)}
    \end{minipage}
\caption{Distribution of polarization emission state on a Poincaré sphere projected using the Hammer equal area projection. Only emissions with the polarization position angle (PPA) error bars less than $5^{\circ}$ for each single pulse were included: (a) the interpulse (left) and main pulse (right); (b) the precursor of the main pulse; (c) the main pulse region (left) and the postcursor of the main pulse (right). The latitudes represent great circular of constant $\chi$ separated by $ 30^{\circ}$ with pure linear polarization being the equator. The north and south poles correspond to pure circular polarization with positive and negative Stokes V, respectively. The pulse number is indicated by the colour bar. The PPAs patches in the $\pm 180^{\circ}$ represent the same polarization mode.}
\label{Poincare}
\end{figure*}

 To unravel the radiation in the pulsar magnetosphere, the polarization emission behaviors of individual pulses are required. As Figure~\ref{Poincare} shows, from the left to right subpanels of the Figure depict the polarization emission state over different emission regions. There are different orthogonal polarization modes (OPMs) that respectively dominate the polarization emission orientation of the interpulse and main pulse (subpanel(a)). It is, however, worth noticing that the polarization emission state in the interpulse is mainly contributed by one type of OPMs, another mode with an ambiguous PPAs patch indicates an extremely weak contribution to the radiation in the pulsar magnetosphere. On the contrary, the polarization emission state in the main pulse exhibits two significant PPAs patches. This polarization feature means that the emission consists of two roughly equal contributions from two modes simultaneously.
 
\par As studied earlier, the emission of the precursor component of the main pulse is unraveled by analyzing the CCF result. The polarization emission state of the precursor component of the main pulse ($90.97^{\circ}$ and $100.12^{\circ}$) is also analyzed to reveal the polarization emission in the magnetosphere. The PPAs patches of the pulse longitude $90.97^{\circ}$ show two distribution properties, while that of the pulse longitude $100.12^{\circ}$ exhibits a line along the equator of the Poincaré sphere. Both polarization emission behaviors indicate that one polarization mode can not describe these phenomena. Moreover, the polarization emission orientation of the pulse longitude $100.12^{\circ}$ becomes more complicated than other pulse longitudes, implying more modes of contributions to the radiation in the magnetosphere.

\par The polarization emission state of the postcursor of the main pulse is not like that of the precursor, showing two modes of contributions. As subpanel (c) shows, the distribution of the single pulse's PPAs exhibits two patches. However, the distribution property of the pulse longitude $136.72^{\circ}$ implies that only one mode dominates the polarization emission in the magnetosphere, for the precursor's case. This polarization behavior is similar to that of the interpulse.

%
%

%
\section{Conclusions and Discussions}
\par The physics of pulsar surface, to be crucial for the radiative mechanism of coherent radio emission, is closely related to the nature of cold matter at supra-nuclear density, which remains still a puzzle in the new era of multi-messenger astronomy~\citep{2021AN....342..320X,2023AN....34430008X}.
Certainly, the stellar surface is hard to probe due to the absence of direct and clear observational evidence.
According to the magnetospheric geometry of this pulsar~\cite{2023arXiv230807691W}, we analyze the single pulses in order to unravel the radiation characteristics and the stellar surface of this pulsar even further.
It is then evident that the non-symmetrical sparking may hint ``zits'' on a pulsar surface~\citep{2023arXiv231205510X}.

\par As discussed earlier for the underlying physical mechanism in different emission regions, the physical processes in the strong emission regions (i.e., the interpulse and main pulse) are similar to each other. A similar physical mechanism occurs between the precursor and postcursor of the main pulse again.
Moreover, the emission feature between the precursor component of the main pulse and the interpulse exhibits a negative correlation. This feature occurs also between the precursor component of the main pulse and the main pulse, showing that another underlying physical process may affect these regions.

\par 
More details about the sparks would be leaked by the emission state of the individual pulses. 
Figure~\ref{Fp} shows that more peak structures are visible in some periods. It could be evidence of the distribution of small mountains on the stellar surface.
For a fixed rotation period, more peak structures could be related to the heights of the mountains. Furthermore, the preferential discharge of the sparks would be related to the heights of the mountains.
In addition, more peak structures could also reflect the distribution between these small mountains.
This emission signature is similar to other periods, implying that the pulsar has ``zits'' on the surface of its polar cap~\citep[e.g.,][]{2023arXiv231205510X,2023arXiv230807691W}.

\par Based on the 110-min polarization observation, we plot the sparking pattern of the polar cap surface of this pulsar. As Figure~\ref{mapping} shows, the sparks are distributed asymmetrically around its magnetic pole.
This feature may indicate that some mountains distribute at the polar cap surface.
This finding is consistent with the observation result of the single pulse shown in Figure~\ref{Fp}.

\par The preferential discharge of the sparks would be closely related to the heights of the mountains. The radio emission peak would correspond to the highest preferential discharge. For most of pulsars, the radio emission peak is aligned with the fiducial plane decided by the magnetic axis and the self-rotation axis. However, the point of highest PA slope of PSR B0950$+$08 is largely misaligned with the radio emission peak (see Figure~\ref{Fp}). Both the emission state of the single pulse and the sparking pattern support that some mountains with different heights are located on the polar cap surface (i.e., there are some “zits” on a pulsar surface).
\par Small mountains on the polar cap can result in a rough surface. As ~\citet{2019SCPMA..6259505L} already reported the emission state of the single pulse of PSR B2016$+$08. They found that the drifting sub-pulses of this pulsar are irregular and diffuse, and argued that these drifting phenomena are due to the rough surface. Furthermore, the rough surface can result in a high-tension point on the stellar surface. A high-tension point discharge may also result in triggering a large bunch of electron-positron pairs ($e^{\pm}$) during a quake-induced oscillation-driven magnetospheric activity for the cases of the soft-gamma-ray repeaters~\citep[e.g., ][]{2015ApJ...799..152L,2006MNRAS.373L..85X}. Similar trigger mechanisms have already been discussed for the central engine of fast radio bursts~\citep[e.g., ][]{2022ApJ...927..105W,2022SCPMA..6589511W}.

\par The polarization emission state of individual pulses is worth studying, which probes the pulsar electrodynamics~\citep[e.g.,][]{1969ApJ...157..869G}. Figure~\ref{Poincare} depicts the polarization emission orientation, showing that more polarization modes dominate the polarization emission behaviors.
Note that the emission feature between the precursor and postcursor of the main pulse exhibits a positive correlation (see Figure~\ref{ccf}), while the polarization emission orientation of these emission regions displays different properties.
These radiation characteristics would imply that different underlying physical mechanisms dominate the emission feature and polarization emission state of this pulsar.
Meanwhile, the contribution of more polarization modes to the radiation in the magnetosphere can not be described by the framework of the RVM geometry.
\par It is worth noticing that the PPA patches are almost lying at the equator, showing that the circular polarization emission contribution to the magnetosphere is extremely weak compared to the linear polarization.
The contribution of the circular polarization in the main pulse becomes significant compared to other pulse longitudes, but it is still a weak contribution compared to the linear polarization in the main pulse.
The underlying physical origins of the polarization mode are not well understood yet. More polarization observations would be required to further understand this issue. 

\par In this work, we further investigate the radiation mechanism of individual pulses of PSR B0950$+$08. To further clarify the underlying physical mechanism over different emission regions, we analyze the cross-correlation function (CCF) map of this pulsar. As Figure~\ref{ccf} shows, different underlying physical mechanisms dominate the emission features. These labeled `` Area 1'', ``Area 2'', ``Area 3'', and ``Area 4'' correspond to the correlation of the emission between different emission regions.

\par As Figure~\ref{Fp} shows, the emission state of this pulsar exhibits more peak structures in some periods. The underlying physical origins of these peak structures can be interpreted as the height of the small mountain on the polar cap surface. Meanwhile, the radio signal of a certain period displays an extremely wide pulse longitude.
For the fixed range of the pulse longitude (e.g., from $100^{\circ}$ to $150^{\circ}$), the micro-structure with different shapes is also detected.
In addition, the fluctuation in emission intensity becomes significant in the main pulse, and this fluctuation occurs over a ms-time scale and becomes frequent in the strong emission region. We are not sure if the emission intensity or time scale of this fluctuation is significantly different from the phenomenon reported by~\cite{2022A&A...658A.143B}.

\par The distribution of the polarization emission orientation of this pulsar is unraveled by the Poincaré sphere view. As Figure~\ref{Poincare} shows, different polarization modes contribute to the radiation in this pulsar magnetosphere. Significant jump feature occurs in the range from the precursor to the postcursor of the main pulse.
Although other PPAs patches are ambiguous in the interpulse, it is evident that two polarization modes dominate the polarization emission state of the interpulse.
\par The contribution of the polarization modes to the magnetosphere exhibits similar features in both the interpulse and the postcursor component of the main pulse, showing that one mode has an overwhelming contribution compared to another.
Meanwhile, the polarization emission state of the precursor component of the main pulse (i.e., $100.12^{\circ}$ of pulse longitude) exhibits a line along the equator. This phenomenon means that the process of the radiation in the magnetosphere over precursor component regions of the main pulse becomes extremely complicated.

\section*{Acknowledgments}
The authors would like to thank those involved in the continuous discussions in the pulsar group at Peking University. This work is supported by the National SKA Program of China (2020SKA0120100), the National Natural Science Foundation of China (12003047 and 12133003), and the Strategic Priority Research Program of the Chinese Academy of Sciences (XDB0550300).
\appendix

\section{The mapping of the spark pattern}

For a static oblique dipole field, the magnetic field,
\begin{equation}
    \mathbf{B} = \frac{\mu}{r^3} \left[ 3 (\mu \cdot \hat{r}) \hat{r} - \mu \right],
\end{equation}
where $\mu$ corresponds to the magnetic moment vector and the $\hat{r}$ is the radial unit vector.

\par Three components of the magnetic field in the Cartesian coordinates is given as follows,
\begin{equation}
    \mathbf{B} = B_{x}^{sd} \hat{x} + B_{y}^{sd} \hat{y} + B_{z}^{sd} \hat{z},
\end{equation}
where 
\begin{equation}
    B_{x}^{sd} = \frac{\mu}{r^5} \left[ 3 x z \cos \alpha + (3x^2 - r^2) \sin \alpha \right],
\end{equation}
\begin{equation}
    B_{y}^{sd} = \frac{\mu}{r^5} \left[ 3yz \cos \alpha + 3xy \sin \alpha \right],
\end{equation}
\begin{equation}
    B_{z}^{sd} = \frac{\mu}{r^5} \left[ (3z^2 - r^2) \cos \alpha + 3 x z \sin \alpha \right],   
\end{equation}
with $\alpha$ the inclination angle.

\par We searched for the solutions by using the Runge-Kutta integration. Firstly, we assume the emission of this pulsar from the magnetic field lines whose footprints concentrate on the three-quarter distance between the magnetic pole and the footprints of the last closed field lines (see Figure~\ref{mapping}). Secondly, according to the relation between the emission location and the viewing angle $\zeta$ given in Equation(~\ref{eq1}), we use the Runge-Kutta integration to follow each field line in space and determine where the direction of the emission point is parallel to the light of sight inside the light cylinder. Finally, the emission point of the field line where the footprint is concentrated on the orange curve is determined.
\bibliography{ref}%

\end{document}